\documentclass[aps,prl,twocolumn,superscriptaddress]{revtex4-2}

\usepackage{amsthm}
\usepackage{amsmath,bm}
\usepackage{amssymb}
\usepackage{amsfonts}
\usepackage{graphicx}
\usepackage{txfonts}
\usepackage{xcolor}
\usepackage{float}
\usepackage{braket}

\usepackage[colorlinks=true,linkcolor=blue,citecolor=blue,urlcolor=blue]{hyperref}

\begin{document}

\title{Distributed quantum sensing with measurement-after-interaction strategies}

\author{Jiajie Guo}
\affiliation{State Key Laboratory for Mesoscopic Physics, School of Physics, Frontiers Science Center for Nano-optoelectronics, $\&$ Collaborative
Innovation Center of Quantum Matter, Peking University, Beijing 100871, China}

\author{Shuheng Liu}
\affiliation{State Key Laboratory for Mesoscopic Physics, School of Physics, Frontiers Science Center for Nano-optoelectronics, $\&$ Collaborative
Innovation Center of Quantum Matter, Peking University, Beijing 100871, China}

\author{Matteo Fadel}
\email{fadelm@phys.ethz.ch}
\affiliation{Department of Physics, ETH Z\"{urich}, 8093 Z\"{urich}, Switzerland}

\author{Qiongyi He}
\email{qiongyihe@pku.edu.cn}
\affiliation{State Key Laboratory for Mesoscopic Physics, School of Physics, Frontiers Science Center for Nano-optoelectronics, $\&$ Collaborative
Innovation Center of Quantum Matter, Peking University, Beijing 100871, China}
\affiliation{Collaborative Innovation Center of Extreme Optics, Shanxi University, Taiyuan, Shanxi 030006, China}
\affiliation{Peking University Yangtze Delta Institute of Optoelectronics, Nantong 226010, Jiangsu, China}
\affiliation{Hefei National Laboratory, Hefei 230088, China}

\begin{abstract}
We investigate multiparameter quantum estimation protocols based on measurement-after-interaction (MAI) strategies, in which the probe state undergoes an additional evolution prior to linear measurements. As we show in our study, this extra evolution enables different level of advantages depending on whether it is implemented locally or nonlocally across the sensing nodes.
By benchmarking MAI strategies in both discrete- and continuous-variable systems, we show that they can significantly enhance multiparameter sensitivity and robustness against detection noise, particularly when non-Gaussian probe states are employed, cases where standard linear measurements are often insufficient.
We also derive analytical results for multiparameter squeezing and establish the corresponding scaling laws for spin-squeezed states, demonstrating that MAI protocols can reach the Heisenberg scaling. These results pave the way for immediate experimental implementation in platforms such as atomic ensembles and optical fields.
\end{abstract}

\maketitle

Distributed quantum sensing aims at simultaneously estimate multiple parameters locally encoded in different nodes~\cite{ZhangQST2021,ProctorPRL2018,GePRL2018}.
Recently, it has been attracting increasing interest due to its relevance to quantum networks~\cite{ProctorPRL2018,ManuelNC2020,YangPRL2024,GePRL2018} and promising applications in a variety of technological applications, such as international clock synchronization~\cite{KomarNP2014} and local beam tracking~\cite{QiarXiv2018}. 
These perspectives lead to intensive investigations on how to enhance the multiparameter sensitivity beyond the classical limit: Various theoretical efforts focus on optimizing the use of entanglement and measurements to achieve the maximum allowed multiparameter sensitivity~\cite{ManuelPRL2018,GePRL2018,YangPRL2024,AuthonyarXiv2024}. In addition, quantum-enhanced multiparameter sensitivity have been demonstrated in both continuous- and discrete-variable experiments~\cite{LiuNP2021,MaliaNature2022,XiaPRL2020,KimNC2024,GuoNP2020,ZhaoPRX2021}. 

\begin{figure}[t]
    \begin{center}
	\includegraphics[width=80mm]{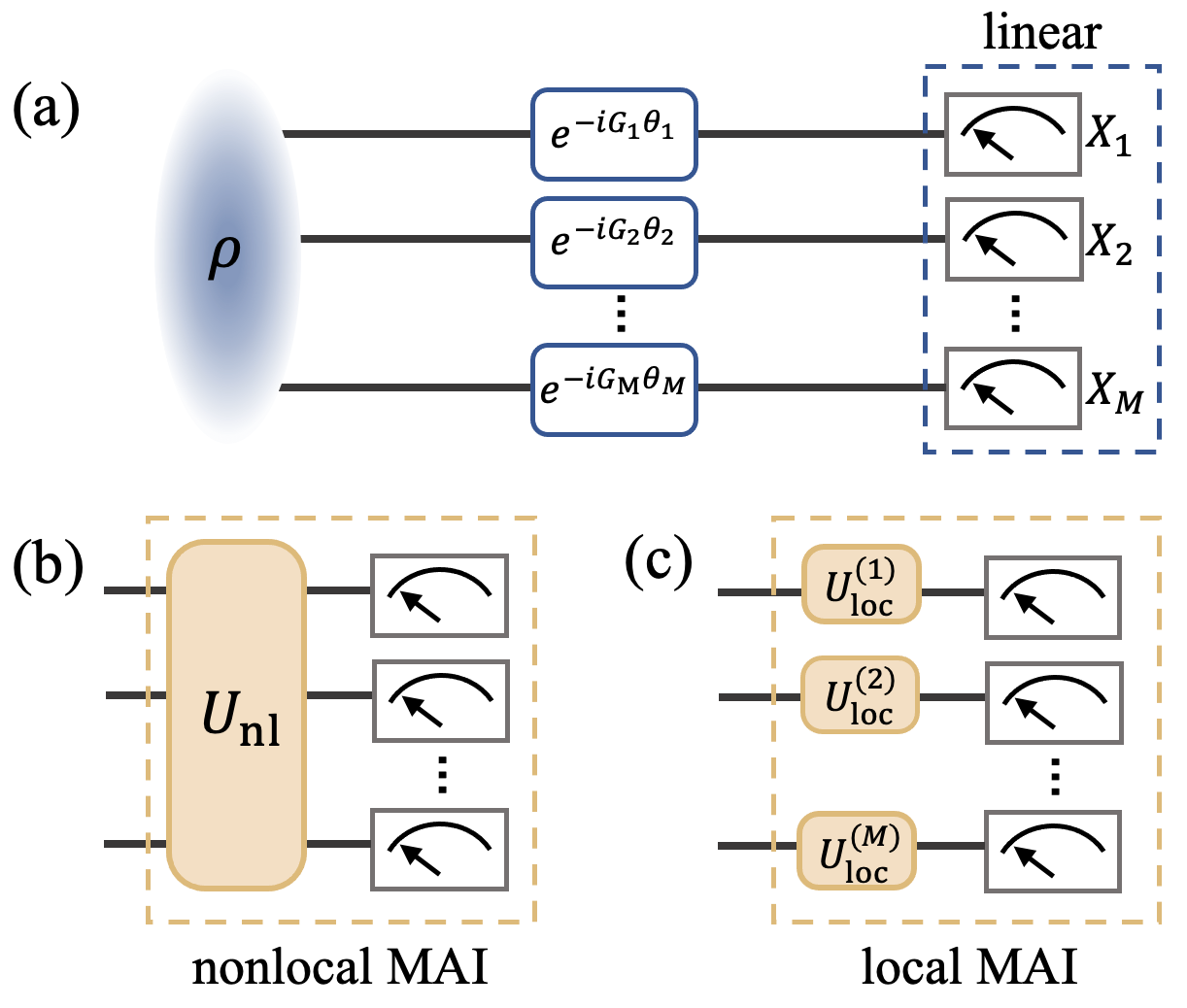}
	\end{center}
    \caption{ Illustration of distributed quantum sensing under different measurement strategies. A set of multiple phases $\boldsymbol{\theta}=\left(\theta_1,\cdots, \theta_M \right)$ are locally encoded to the probe state $\rho$, where $G_m$ are commuting phase generators. Then the multiparameter sensitivity will be estimated based on the results from (a) a set of local linear measurements $X_m$. In the MAI protocol, (b) nonlocal evolutions $U_{\text{nl}}$ or (c) local evolutions $U^{(m)}_{\text{loc}}$ are introduced before the linear measurements.}
    \label{Fig1_Illustration}
\end{figure}

In typical multiparameter estimation protocols, especially the one involving atomic ensembles or optical fields, linear measurements are performed on the probe states in order to estimate the unknown parameters~\cite{ManuelNC2020,MaliaNature2022,GuoNP2020,MatteoNJP2023}, see Fig.~\ref{Fig1_Illustration}(a).
However, this measurement strategy, although experimentally easy to implement, comes at the cost of reduced detection capability. 
Linear measurements are, in fact, insensitive to non-Gaussian properties of the probe states, which could provide a significant additional advantage in the metrological task.
For this reason, taking inspiration from single parameter estimation scenarios, we introduce the measurement-after-interaction (MAI) strategy to distributed sensing tasks.
This approach consists of adding an additional state evolution before the linear measurements to achieve an increased signal-to-noise ratio~\cite{DavisPRL2016,FrowisPRL2016,TommasoPRA2016,NolanPRL2017,MariusQuantum2020,YoucefPRL2021,JiajiePRA2024}. 
Without additional measurement statistics, the MAI strategy leads to enhanced noise robustness as well as in revealing the high sensitivity of non-Gaussian probe states, as demonstrated in remarkable experiments~\cite{HostenScience2016,BurdScience2019,SimoneNP2022,QiNP2022}.

When it comes to the framework of multiparameter estimation, we propose and compare two complementary scenarios: nonlocal and local MAI strategies. 
In the case of nonlocal MAI, see Fig.~\ref{Fig1_Illustration}(b), the additional state evolution is implemented on the whole probe state. 
However, since in many distributed sensing tasks the individual nodes can be far apart, a more practical strategy could be a local MAI protocol, see Fig.~\ref{Fig1_Illustration}(c), where the additional evolution is individually implemented on each node.
The question is on how these protocols compare in terms of the achievable multiparameter estimation sensitivity and noise robustness. 

In this work, we provide a general framework to estimate multiparameter sensitivity using MAI strategies. We benchmark both nonlocal and local protocols in both discrete- and continuous-variable systems, and demonstrate that both protocols show enhanced detection capability compared to the typical scenario solely relying on local linear measurements. 
To consider experimentally relevant situations, we first analyze the paradigmatic case of multi-mode spin-squeezed states generated from the one-axis-twisting interaction, where the multiparameter sensitivity is determined by both mode entanglement and particle entanglement~\cite{ManuelPRL2018}. We show that the nonlocal MAI strategy is more powerful to capture mode entanglement compared to its local counterpart. Furthermore, we analytically derive the scaling laws under different measurement strategies, and surprisingly find that besides nonlocal MAI protocol, the local MAI protocol with small mode numbers is able to achieve the Heisenberg scaling. In addition, we further investigate two-mode squeezed vacuum states, as a simple case of continuous-variable Gaussian states, and show that the robustness against detection noises is significantly improved in MAI strategies.

\vspace{2mm}
\textbf{Distributed quantum sensing and the multiparameter squeezing matrix.--}
The distributed sensing scenario can be described by a multimode interferometer, see Fig.~\ref{Fig1_Illustration}, where in each of the $M$ paths it is encoded an unknown phase $\theta_m$. 
The goal is then to estimate an arbitrary linear combination of these phases, $\Theta = \mathbf{n}^T \boldsymbol{\theta}$, where $\mathbf{n}=(n_1,\cdots,n_M)$, $|\mathbf{n}|=1$, specifies the linear combination and $\boldsymbol{\theta}=\left(\theta_1,\cdots,\theta_M \right)$. 
These parameters are encoded on the probe state via the unitary evolution $\rho(\boldsymbol{\theta})=e^{-i \mathbf{G} \boldsymbol{\theta}} \rho e^{i \mathbf{G} \boldsymbol{\theta}}$, where $\mathbf{G}=\left( G_1,\cdots,G_M \right)$ is a vector of local generators that commute with each other.
After a sequence of $\nu$ independent measurements performed on the probe $\rho(\boldsymbol{\theta})$, an unbiased estimator $\boldsymbol{\theta}_{\text{est}}=(\theta_{\text{est},1},\cdots,\theta_{\text{est},M})$ for $\boldsymbol{\theta}$ can be constructed from the readouts. 
To represent the error of the estimation, a covariance matrix $\boldsymbol{\Sigma}$ with elements $\boldsymbol{\Sigma}_{ij}=\text{Cov}(\theta_{\text{est},i},\theta_{\text{est},j})$ is defined, which yields the variance of the linear combination of multiparameter estimation $\text{Var}[\Theta_{\text{est}}]= \mathbf{n}^T \boldsymbol{\Sigma} \mathbf{n}$ with $\Theta_{\text{est}} = \mathbf{n}^T\boldsymbol{\theta}_{\text{est}}$. 
Due to the fundamental limits given by the Cram$\acute{\text{e}}$r-Rao and quantum Cram$\acute{\text{e}}$r-Rao bounds~\cite{HelstromJSP1969,Kaybook1993,BraunsteinPRL1994}, it holds the hierarchy of sensitivities $\boldsymbol{\Sigma} \geq \mathbf{F}^{-1}/\nu \geq   \mathbf{F}_Q^{-1}/\nu $, where $\mathbf{F}$ and $\mathbf{F}_Q$ are classical and quantum Fisher information matrix, respectively.

One of the commonly used estimator is known as the method of moments, which consists of estimating the unknown parameter from a moment of the measurement probability distribution. Let us consider a family of accessible local measurement operators $\mathbf{X}=\left(X_1,\cdots,X_M \right)$, from which we express the moment-based covariance matrix as 
\begin{align}\label{eq:SigmaCM}
\boldsymbol{\Sigma} = \left( \nu \mathbf{M}[\rho, \mathbf{G},\mathbf{X}] \right)^{-1},
\end{align}
with the moment matrix
\begin{align}\label{eq:momentmatrix}
\mathbf{M}[\rho, \mathbf{G},\mathbf{X}] = \mathbf{C}[\rho, \mathbf{G},\mathbf{X}]^T \boldsymbol{\Gamma}[\rho,\mathbf{X}]^{-1} \mathbf{C}[\rho, \mathbf{G},\mathbf{X}].
\end{align}
Here, $ \boldsymbol{\Gamma}[\rho,\mathbf{X}]$ is the covariance matrix with elements $\left(\boldsymbol{\Gamma}[\rho,\mathbf{X}] \right)_{ij}=\text{Cov}(X_i,X_j)$, and $\mathbf{C}[\rho, \mathbf{G},\mathbf{X}]$ is the commutator matrix with elements $\left(\mathbf{C}[\rho, \mathbf{G},\mathbf{X}]\right)_{ij}=-i\langle [ X_i, G_j ]\rangle$. 

The shot-noise limit is given as $\boldsymbol{\Sigma}_{\text{SN}}=\left( \nu \mathbf{F}_{\text{SN}}[\mathbf{G}] \right)^{-1}$, where $\mathbf{F}_{\text{SN}}$ is the quantum Fisher information matrix determining the sensitivity limit from a classical scheme. To quantify the quantum-enhanced sensitivity above such classical limits, we make use of the squeezing matrix
\begin{align}\label{eq:squeezingmatrix}
\Xi^2[\rho,\mathbf{G},\mathbf{X}]= \mathbf{F}_{\text{SN}}[\mathbf{G}]^{\frac{1}{2}} \mathbf{M}[\rho,\mathbf{G},\mathbf{X}]^{-1} \mathbf{F}_{\text{SN}} [\mathbf{G}]^{\frac{1}{2}}.
\end{align}
Observing $\Xi^2[\rho,\mathbf{G},\mathbf{X}] < \mathbb{I}_{M}$ reveals a quantum-enhanced sensitivity in the multiparameter scenario, i.e. $\boldsymbol{\Sigma}< \boldsymbol{\Sigma}_{\text{SN}}$.

\vspace{2mm}
\textbf{Nonlocal and local measurement-after-interaction (MAI) protocols.--}
In typical experimental platforms for metrology, the moment-based sensitivity is determined by performing linear measurements, such as collective spin observables in atomic systems or quadrature operators in light fields. Although these linear measurements are optimal to reveal the sensitivity of Gaussian states undergoing Gaussian transformations, they are insufficient for non-Gaussian states. 
The high metrological potential of these states, in fact, resides in high-order moments that are more difficult to access.

An experimentally practical approach to effectively measure these high moments without the need to change detection apparatus consists of the measurement-after-interaction (MAI) protocol~\cite{DavisPRL2016,FrowisPRL2016,TommasoPRA2016,NolanPRL2017}. 
In this protocol, the probe state will go through a unitary evolution $U(\tau)=e^{-i H \tau}$, with $H$ a nontrivial Hamiltonian, before it is probed by linear measurements. 
The MAI protocol has been widely investigated in single-parameter estimation tasks, showing its advantages on improving the robustness against detection noise as well as revealing sensitivity in non-Gaussian states~\cite{HostenScience2016,BurdScience2019,SimoneNP2022,QiNP2022,YoucefPRL2021,JiajiePRA2024}. 
For multi-parameter estimation, numerical studies for an atomic ensemble where each atom carries an independent phase have suggested that MAI strategies can provide an advantage~\cite{YoucefSPP2023}.
In the following, we provide a completely general framework for exploring the application of MAI methods in a distributed sensing scenario, providing analytical results.

First, let us introduce a family of accessible linear operators on the $m$-th mode, $\mathcal{L}=\left( L_{1},\cdots,L_{K} \right)$. 
In the typical scheme as in Fig.~\ref{Fig1_Illustration}(a), both phase generators and measurements are considered to be local linear operators. Therefore, they can be expressed as $G_{m} =G_{\text{L}}^{(m)} := \mathbf{g}_{m}^T \mathcal{L}$ and $X_m=X_{\text{L}}^{(m)} := \mathbf{x}_{m}^T \mathcal{L}$, respectively, where $\mathbf{g}_m,\mathbf{x}_m$ are normalized real vectors. 
Then, based on the specific requirements of the considered estimation task, we investigate two possible measurement strategies: nonlocal MAI and local MAI, which are illustrated in Fig.~\ref{Fig1_Illustration}(b) and Fig.~\ref{Fig1_Illustration}(c), respectively. 

In a nonlocal MAI strategy, a unitary evolution $U_{\text{nl}}=e^{-i H_{\text{nl}} \tau}$ is applied on the whole system before the local measurements are performed. This process is equivalent to performing a nonlocal MAI measurement on $m$-th mode, which reads
\begin{align}
X_{\text{MAI,nl}}^{(m)}  &= U^\dagger_{\text{nl}} X_{\text{L}}^{(m)}   U_{\text{nl}} .
\end{align}
In a local MAI strategy, on the other hand, we consider $M$ unitary evolutions each performed on a single mode, i.e. $U_{\text{loc}}^{(m)} =e^{-i H^{(m)}_{\text{loc}} \tau^{(m)} }$, such that the resulting local MAI operator can be expressed as
\begin{align}
X_{\text{MAI,loc}}^{(m)}  &= U_{\text{loc}}^{(m) \dagger}  X_{\text{L}}^{(m)}  U_{\text{loc}}^{(m) } .
\end{align}
Note that, in general, these local unitary evolutions can be different from each other.

\vspace{2mm}
\textbf{Measurements optimization in MAI strategy.--}
Given a vector of accessible linear operators $\mathcal{L}^{(m)}=\left(L^{(m)}_1,\cdots,L^{(m)}_K \right)$, we can construct the family of linear measurements over all $M$ modes as $\mathcal{A}_{\text{L}}=\left( \mathcal{L}^{(1)},\cdots, \mathcal{L}^{(M)} \right)$. In the MAI strategy, an MAI vector $\mathbf{X}^{(m)}_{\text{MAI}}$ is determined by an additional evolution $U(\tau)$, whose elements can be expressed as $\mathbf{X}^{(m)}_{\text{MAI},k}=U^\dagger (\tau) L^{(m)}_{k} U(\tau)$, so that one can obtain a time-dependent MAI family $\mathcal{A}_{\text{MAI}}(\tau)=\left( \mathbf{X}^{(1)}_{\text{MAI}} (\tau),\cdots, \mathbf{X}^{(M)}_{\text{MAI}} (\tau) \right)$. With these accessible measurements settings at hand, the vectors for phase generator and measurements can be reexpressed as $\mathbf{G}=R\mathcal{A}_{\text{L}}$ and $\mathbf{X}(\tau)= S\mathcal{A}_{\text{MAI}}(\tau)$, where $R$ and $S$ are $M\times (MK)$ real matrices satisfying $RR^T=S S^T=\mathbb{I}_{M} $.

According to the method proposed in~\cite{ManuelNC2020}, for any transformation matrix $R$, the maximum moment matrix in Eq.~\eqref{eq:momentmatrix} is given by
\begin{align}
\max_{\mathbf{X}\in \text{span}(\mathcal{A}_{\text{MAI}}(\tau))} \mathbf{M}[\rho,\mathbf{G},\mathbf{X}(\tau)]= R \mathbf{M} [\rho,\mathcal{A}_{\text{L}},\mathcal{A}_{\text{MAI}}(\tau)] R^T,
\end{align}
where $\mathbf{M} [\rho,\mathcal{A}_{\text{L}},\mathcal{A}_{\text{MAI}}(\tau)]= \mathbf{C}[\rho,\mathcal{A}_{\text{L}},\mathcal{A}_{\text{MAI}}(\tau)]^T $ $\boldsymbol{\Gamma}[\rho,\mathcal{A}_{\text{MAI}}(\tau)]^{-1}   \mathbf{C}[\rho,\mathcal{A}_{\text{L}},\mathcal{A}_{\text{MAI}}(\tau)]$ is the moment matrix. Here, the corresponding optimal measurements matrix holds $GS=R\mathbf{C}[\rho,\mathcal{A}_{\text{L}},\mathcal{A}_{\text{MAI}}(\tau)]^T \boldsymbol{\Gamma}[\rho,\mathcal{A}_{\text{MAI}}(\tau)]^{-1}$, where $G$ is a real-value matrix for normalization. Based on the moment matrix, one can obtain the optimal squeezing matrix in Eq.~\eqref{eq:squeezingmatrix}
\begin{align}
  & \boldsymbol{\Xi}^2_{\text{opt}} [\rho,\mathbf{G},\mathcal{A}_{\text{MAI}}(\tau)] :=\min_{\mathbf{X}(\tau)\in\text{span}(\mathcal{A}_{\text{MAI}}(\tau))}  \boldsymbol{\Xi}^2[\rho,\mathbf{G},\mathbf{X}(\tau)] \nonumber \\
  &= \mathbf{F}_{\text{SN}}^{\frac{1}{2}}[\mathbf{G}] R \mathbf{M}[\rho,\mathcal{A}_{\text{L}},\mathcal{A}_{\text{MAI}}(\tau)]^{-1} R^T \mathbf{F}_{\text{SN}}^{\frac{1}{2}}[\mathbf{G}],
\end{align}
thus the covariance matrix in Eq.~\eqref{eq:SigmaCM} can be reexpressed as
\begin{align}
\Sigma[\rho,\mathbf{G},\mathcal{A}_{\text{MAI}}(\tau)]=\Sigma_{\text{SN}}^{\frac{1}{2}} \boldsymbol{\Xi}^2_{\text{opt}} [\rho,\mathbf{G},\mathcal{A}_{\text{MAI}}(\tau)] \Sigma_{\text{SN}}^{\frac{1}{2}} 
\end{align}
For a given vector $\mathbf{n}$ and an accessible time range $T$ for the MAI unitary evolution, the relative estimation uncertainty of a linear combination of imprinted parameters is given by
\begin{align}
\xi^{-2} (\mathbf{n}) = \max_{\tau \in \text{span} (T) } \frac{\mathbf{n}^T \Sigma_{\text{SN}} \mathbf{n} }{\mathbf{n}^T \Sigma[\rho,\mathbf{G},\mathcal{A}_{\text{MAI}}(\tau)] \mathbf{n}}.
\end{align}
Here, $\xi^{-2} (\mathbf{n})>1$ reveals metrologically useful squeezing.

\begin{figure}[t]
    \begin{center}
	\includegraphics[width=85mm]{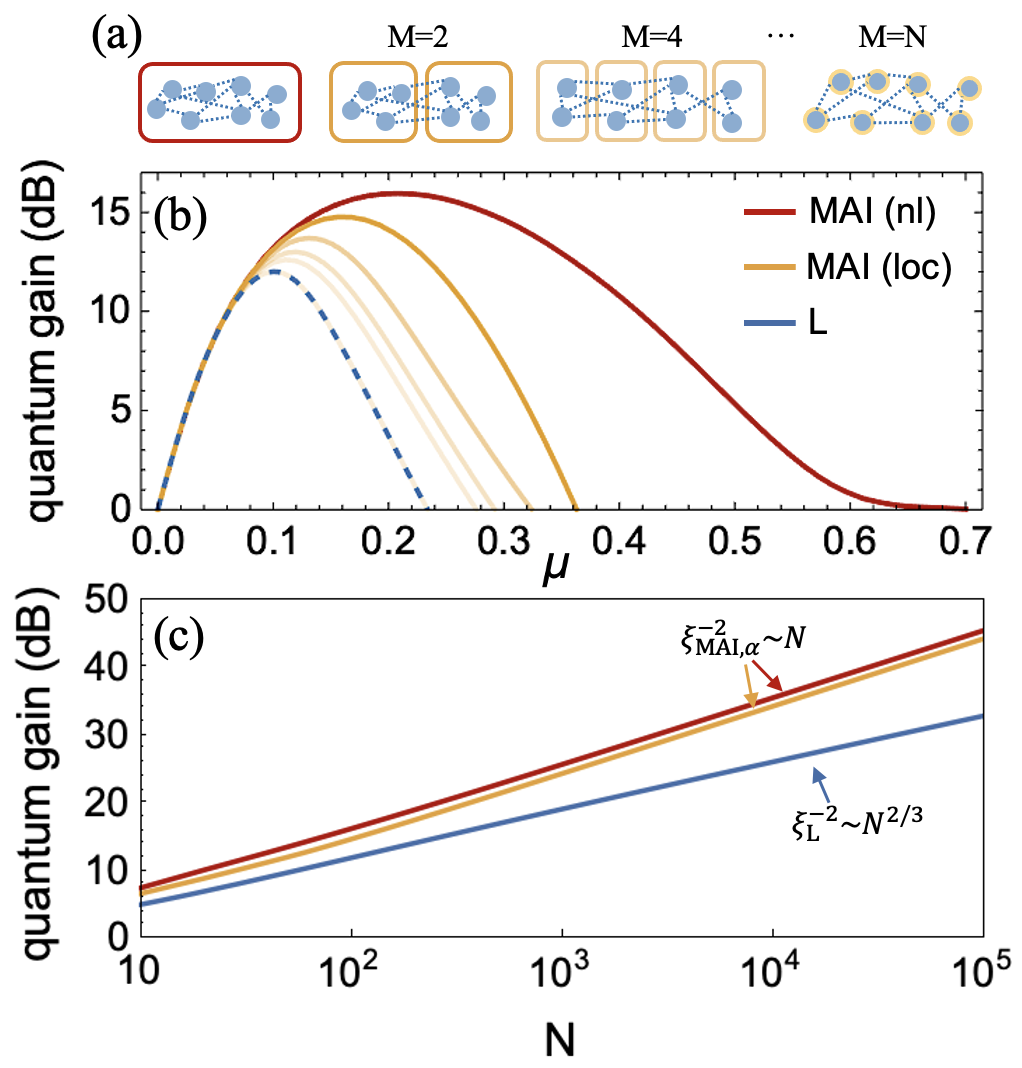}
	\end{center}
    \caption{Multiparameter squeezing for mode-entangled spin squeezed states $\rho_{\text{ME}}$ under different measurement strategies. 
    (a) Illustration of nonlocal and local MAI: The nonlocal MAI is performed on the entire ensemble of $N$ particles (red box), while the local MAI are separately performed on $N_m$ particles within each modes (yellow boxes). The number of local atoms $N_m$ will decrease as $M$ increases, and eventually reaches $N_m=1$ when $M=N$. (b) Quantum gain $10 \text{log}_{10}(\xi^{-2})$ as a function of the evolution time $\mu=2\chi t$, for $N=100$. The sensitivities detected from nonlocal MAI (red solid line) and linear measurements (blue solid line) are independent of $M$. The sensitivity from local MAI will decrease with $M$ (yellow lines), where the opacity decreases with mode number $M$ ($M=2,4,10,20,100$ is shown).
    (b) Sensitivity scaling for linear measurements, nonlocal MAI (independent with $M$) and local MAI ($M=2$) as the function of total atom number $N$. Both MAI protocols yield Heisenberg scaling $\xi^{-2}_{\text{MAI},\alpha} \sim N$, which significantly outperform the one from linear measurement $\xi^{-2}_{\text{L}} \sim N^{2/3}$.  }
    \label{Fig2_spindB}
\end{figure}

\vspace{2mm}
\textbf{Multiparameter squeezing for collective spin system.--}
Let us first benchmark MAI strategies with a paradigmatic case of spin squeezed states in atomic experiments. These states are of immediate practical relevance for atomic ensembles, and have been widely used in quantum metrology to overcome the classical shot-noise limit. Moreover, state-of-the-art atomic experiments~\cite{MatteoScience2018,PhilippScience2018, PaoloPRX2023} have demonstrated that the entanglement among the constituent particles can be preserved even after spin-squeezed states are spatially split into different spatial modes. Such distributed entanglement is found to enable uncertainty reduction in multi-parameter estimation~\cite{ManuelNC2020,MatteoNJP2023}, making split spin-squeezed states a promising resource for distributed quantum sensing.

We consider the scenario where an ensemble of $N$ spin-$1/2$ particles polarized along the $x$ direction are distributed among $M$ spatially-separated modes. Then, the spin states evolve for a time $t$ according to the one-axis twisting (OAT) interaction described by the nonlocal Hamiltonian
\begin{align}
\tilde{H}_{\text{nl}}  &=\hbar \chi \left(\sum_{m=1}^{M} S_{z}^{(m)} \right)^2.
\end{align}
Here, $S_{z}^{(m)} = \sum_{i}^{N_m } \sigma_{z}^{(i)}/2 $ represents the collective spin operators for the $N_m $ atoms in the $m$-th mode and $\chi$ is the interaction strength. The nonlocal operator $\tilde{H}_{\text{nl}}$ simultaneously generates both mode entanglement and particle entanglement~\cite{jing_split_2019,FadelPRA20,MatteoNJP2023}. To understand the metrological advantage resulting from these different types of entanglement, we take as reference the state prepared through the local Hamiltonian
\begin{align}
\tilde{H}_{\text{loc}} &=\hbar \chi \sum_{m=1}^{M} \left(S_{z}^{(m)} \right)^{2}.
\end{align}
This operator can generate particle entanglement locally, although it is incapable of producing mode entanglement. In the following, we shorthand the resulting states from nonlocal and local OAT interactions as mode entangled states $\rho_{\text{ME}}$ and mode separable states $\rho_{\text{MS}}$, respectively.

In typical strategies considering linear measurement observables, since both squeezing and anti-squeezing directions are on the $yz$-plane, one can optimize the generator $\mathbf{G}$ and measurement $\mathbf{X}$ directions over the basis $\mathcal{L}^{(m)}=\left( S_{y}^{(m)},S_{z}^{(m)} \right)$. 
In MAI strategies, on the other hand, an additional evolution $U(\tau)=e^{-i\frac{H}{\hbar}\tau}$ before linear measurements is considered. 
Naturally, we choose the MAI Hamiltonian $H$ to be of the same type of the state-preparation Hamiltonian $\tilde{H}$. 
To further simplify our calculations, we also consider the local atom number and local MAI evolutions in each sub-ensemble to be the same, i.e. $N_m=N/M$ and $U_{\text{loc}}^{(m)}(\tau^{(m)})=U_{\text{loc}}(\tau)$. 
We can then construct the families of MAI measurements as $\mathbf{X}^{(m)}_{\text{MAI},\alpha}= \left( U^\dagger_{\alpha}(\tau) S_{y}^{(m)} U_{\alpha}(\tau) , U^\dagger_{\alpha}(\tau) S_{z}^{(m)} U_{\alpha}(\tau) \right) $, where $U_{\text{loc}}(\tau)=e^{-i \chi \tau \left(S_{z}^{(m)} \right)^2 }$ and $U_{\text{nl}}(\tau)=e^{-i \chi \tau \left( \sum_m S_{z}^{(m)} \right)^2 }$ are the unitary operators in the local and nonlocal MAI strategies, respectively. 
In the metrological task we are investigating, our aim is to estimate the linear combination $\Theta=\mathbf{n}^T \boldsymbol{\theta}=\sum_{m=1}^M n_m \theta_m$, where $\mathbf{n}=(n_1,\cdots,n_M)$ is a vectors of equal weights with different signs, $n_m=\pm 1 / \sqrt{M}$. Since commuting phase generators are considered, the maximum sensitivity after measurement optimization is independent of the signs in $\mathbf{n}$, that is $\xi^{-2}(\mathbf{n})=\xi^{-2}$.

We present in Fig.~\ref{Fig2_spindB}(b) the multiparameter sensitivity for mode-entangled spin-squeezed states $\rho_{\text{ME}}$ as a function of OAT evolution time $\mu=2\chi t$. 
At short evolution times, OAT spin states are nearly Gaussian, thus both linear and MAI strategies can detect their metrological usefulness. 
However, as $\mu$ increases the states will become over-squeezed and thus non-Gaussian. 
For this reason, in such a regime the sensitivity revealed by linear measurements decay fast. 
MAI strategies, on the contrary, keep showing quantum-enhanced sensitivity for a wider time-evolution range, which is due to the additional unitary before measurement. 
Based on the analytical expressions of moment matrices in SM Sec. II, we find that the resulting sensitivity of nonlocal MAI and linear measurements is independent with mode number $M$. 
However, this is not the case for local MAI. When $M$ is small, the local MAI strategy can achieve optimal sensitivity comparable to that of the nonlocal MAI strategy. While, as $M$ increases, the sensitivity of local MAI decreases and eventually matches that of linear measurements in the extreme case of $M=N$. 
The reason behind this being that for individual spin-1/2 particles the OAT Hamiltonian acts trivially.
In addition, we compare the multiparameter sensitivity for mode separable states $\rho_{\text{MS}}$ in SM Sec. II, to verify the contribution of mode entanglement to the metrological advantage~\cite{ManuelPRL2018,ManuelNC2020,MatteoNJP2023}.

The scaling laws of multiparameter sensitivity for $\rho_{\text{ME}}$ as a function of total atom number $N$ are shown in Fig.~\ref{Fig2_spindB}(c), and the details of analytical solutions for associated moment matrix are given in SM Sec. II. 
For the experimentally practical case of short times scales and relatively large atom number $N$, we observe that the multiparameter sensitivity for the nonlocal MAI protocol can reach the Heisenberg scaling $\xi^{-2}_{\text{MAI,nl}} \sim N$  (red line) when the optimal MAI evolution corresponds to a time-reversal evolution $\tau_{\text{opt}} \approx t$, and it shows a significant advantage over the one from linear measurements $\xi^{-2}_{\text{L}} \sim N^{2/3}$ (blue line). 
In addition, although the moment matrices in linear measurements and nonlocal MAI strategies are determined by the mode number $M$ (see SM Sec. II), the maximum multiparameter squeezing coefficients yielded for both strategies are independent of $M$. 
Furthermore, in Fig.~\ref{Fig2_spindB}(c) we also plot the sensitivity scaling for the local MAI protocol in the specific case of $M=2$ (yellow line). Under the optimal MAI evolution times $\tau_{\text{opt}} \approx 2t$, the local MAI strategy exhibits a metrological advantage comparable to the nonlocal MAI, and is also able to reach the Heisenberg scaling $\xi^{-2}_{\text{MAI,loc}} \sim N$.

\begin{figure}[t]
    \begin{center}
	\includegraphics[width=85mm]{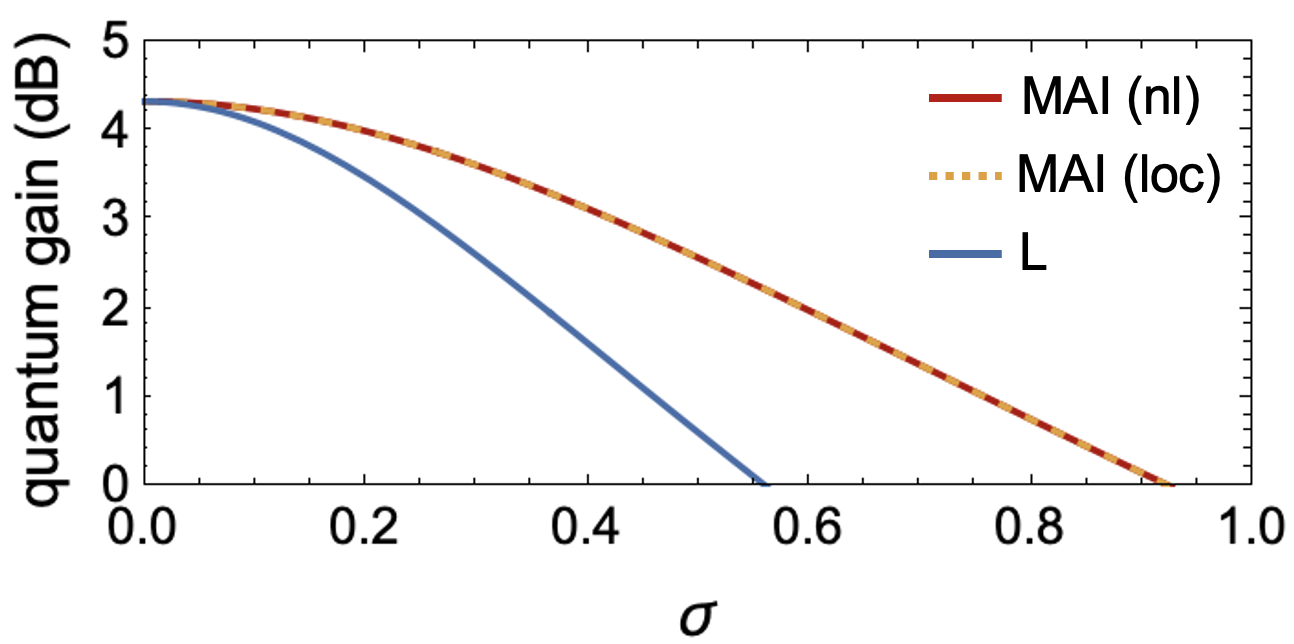}
	\end{center}
    \caption{Detection noise effects on the multiparameter sensitivity. For a two-mode squeezed vacuum state with squeezing parameter $r=0.5$, we show the quantum gain obtained from linear measurements and MAI strategies ($r_{\text{nl}}=r_{\text{loc}}=r$) as a function of standard derivation $\sigma$ of Gaussian-distributed detection noise. }
    \label{Fig3_CVsqueezed}
\end{figure}

\vspace{2mm}
\textbf{Multiparameter squeezing for continuous-variable system.--}
As a simple, yet immediately practical, example in continuous-variable (CV) systems, we investigate the advantage of implementing the local/nonlocal MAI strategies on a two-mode squeezed vacuum (TMSV) state. 
Such states are routinely prepared in various CV platforms, and allow for quantum-enhanced (multi)parameter estimation tasks in SU(1,1)-type interferometers~\cite{OuAPL2020}.
Mathematically, TMSV are described by the action of the two-mode squeezing operator $S_{AB}=\text{exp}\left[-\zeta a^\dagger b^\dagger +\zeta^* a b \right]$ on the vacuum state, where $\zeta=re^{i\phi}$ is the squeezing parameter. 
The family of linear measurements contains quadrature operators $\mathcal{L}=\left( x,p\right)$ with $x=(a+a^\dagger )/\sqrt{2}, p=-i(a-a^\dagger)/\sqrt{2}$. 
Then, the additional evolutions in nonlocal and local MAI strategies can be considered as a two-mode squeezing operator $U_{\text{nl}}=\text{exp}\left[ \zeta_{\text{nl}} a^\dagger b^\dagger - \zeta_{\text{nl}}^* ab \right]$ and one-mode squeezing operators $U_{\text{loc}}=\text{exp} \left[ (\zeta_{\text{loc}}(a^{\dagger})^2-\zeta_{\text{loc}}^* a^2 )/2 \right]$ with $\zeta_\alpha = r_\alpha e^{i\phi_\alpha} $, respectively. 
In a multimode displacements estimation task with Gaussian states, the linear measurements are optimal to saturate quantum Cram$\acute{\text{e}}$r-Rao bound. Therefore, we find that the linear measurements and MAI strategies will yield the same sensitivity, $\xi^{-2}_{\text{L}}=\xi^{-2}_{\text{MAI},\alpha}=e^{2r}$, which indicates the reveal of metrological entanglement if $r>0$ (see SM Sec. III). In such ideal scenario, the sensitivity is independent of MAI squeezing $r_{\alpha}$, indicating that the MAI schemes do not offer any enhancement in metrological advantage.

However, when the effects of detection noise are taken into account, we show that MAI schemes will exhibit enhanced noise robustness. Consider Gaussian-distributed detection noises with standard deviation $\sigma$ (see SM Sec. III), the ratio of the sensitivities from typical and MAI strategies is given as 
\begin{align}
\frac{\xi^{-2}_{\text{L}}}{\xi^{-2}_{\text{MAI},\alpha}} &= \frac{e^{-2r}+2\sigma^2 e^{-2r_{\alpha}} }{ e^{-2r}+2\sigma^2 },
\end{align}
demonstrating that MAI strategies can outperform linear measurements if $r_{\alpha} >0$. In Fig.~\ref{Fig3_CVsqueezed}, we present the quantum gain as a function of squeezing level $r$, for $r_{\text{loc}}= r_{\text{nl}}=r$. Both MAI schemes show significant advantage on noise robustness over the typical scheme.

\vspace{2mm}
\textbf{Conclusions.--} 
We have characterized the performances of MAI strategies in distributed quantum sensing, showing in which cases they allow for an enhancement in multiparameter sensitivity. 
In the considered MAI strategies, an additional evolution, which can be either nonlocal or local, is implemented before a linear measurement. 
Benchmarking with both discrete- and continuous-variable systems, we show that MAI strategies significantly outperform the typical scenario in detection of non-Gaussian sensitivity as well as robusteness against detection noise. Furthermore, we give the analytical solutions for multiparameter squeezing for the spin states generated by the one-axis-twisting (OAT) interaction, and reveal the fundamental principles of multiparameter estimation under different measurement strategies. Surprisingly, we show that the Heisenberg scalings can be achieved in both MAI protocols. 

In our framework of MAI strategies, the evolution is not limited to specific Hamiltonians. If a wider class of experimentally accessible interactions is available, the additional evolution can be further designed and optimized, e.g. by using reinforcement learning~\cite{CaoPRL2023}, to reach a higher sensitivity. 
Similar ideas could be explored to use MAI strategies for the detection of non-Gaussian quantum correlations \cite{KitzingerPRA21}.
Besides providing timely and powerful tools in multiparameter estimation scenarios, our work expands the metrological applications of non-Gaussian states, and provide practical guidelines for implementing distributed sensing tasks in the state-of-the-art experiments.

\vspace{2mm}
\textit{Acknowledgments.--} We thank Youcef Baamara and Alice Sinatra for helpful advice. This work was supported by the National Natural Science Foundation of China (No. 12125402, No. 12447157, No. 12405005), Beijing Natural Science Foundation (Grant No. Z240007), and the Innovation Program for Quantum Science and Technology (No. 2024ZD0302401). J.G. acknowledges Postdoctoral Fellowship Program of CPSF (GZB20240027), and the China Postdoctoral Science Foundation (No. 2024M760072). S.L. acknowledges the China Postdoctoral Science Foundation (No. 2023M740119). M.F. was supported by the Swiss National Science Foundation Ambizione Grant No. 208886, and by The Branco Weiss Fellowship -- Society in Science, administered by the ETH Z\"{u}rich.

\bibliographystyle{apsrev4-1} 
\bibliography{Reference}

\clearpage
\newpage

\begin{widetext}

\section{Supplemental material for Distributed quantum sensing with measurement-after-interaction strategies}

\section{I.\quad Measurement optimization for multiparameter squeezing}
We use the methods of moment in \cite{ManuelNC2020} to analytically optimize the measurements among a set of accessible operators in the MAI protocols. Let us first consider a family of linear measurements $\mathcal{A}_{\text{L}}=\left( \mathcal{L}^{(1)},\cdots, \mathcal{L}^{(M)} \right)$, where $\mathcal{L}^{(m)}=\left( L^{(m)}_1, \cdots, L^{(m)}_K\right)$ represent the vectors of accessible linear operators on the $m$-th mode. In linear-encoding scenarios, a local phase generator can be expressed by  $G_m=\mathbf{g}_m^T \mathcal{L}^{(m)}=\sum_{k=1}^K g_{m,k} L_k^{(m)}$, such that the family consisting of $M$ commuting generators is given as $\mathbf{G}=\left( G_1,\cdots, G_M \right) = R \mathcal{A}_{\text{L}}$, where
\begin{align}
R = \left( \begin{matrix} 
\mathbf{g}_1 & \cdots & \mathbf{0} \\  
\vdots & \ddots & \vdots \\
\mathbf{0} & \cdots &  \mathbf{g}_M
\end{matrix} \right)
\end{align}
is an $M\times (MK)$ real matrix satisfying $RR^T=\mathbb{I}_{M}$. Then, given the unitary evolutions for MAI technique, $U_\alpha=e^{-i \frac{H_\alpha}{\hbar} \tau}$ with $\alpha=\{ \text{nl},\text{loc} \}$, one can construct the vectors of measurements on $m$-th mode, $\mathbf{X}_{\text{MAI},\alpha}^{(m)}(\tau)=\left( U_\alpha^\dagger (\tau) L^{(m)}_1 U_\alpha(\tau),\cdots, U_\alpha^\dagger (\tau) L^{(m)}_K U_\alpha(\tau) \right)$, and further obtain a time-dependent MAI vector on the entire $M$-mode system  $\mathcal{A}_{\text{MAI}}(\tau)=\left( \mathbf{X}_{\text{MAI}}^{(1)}(\tau),\cdots,\mathbf{X}_{\text{MAI}}^{(M)}(\tau) \right)$. Based on such family, any measurements can thus be given as $\mathbf{X}(\tau) = S \mathcal{A}_{\text{MAI}}(\tau)$, where $S$ is an $M\times (MK)$ real matrix satisfying $SS^T=\mathbb{I}_{M}$. 

Using $\mathbf{C}[\rho,\mathbf{G},\mathbf{X}(\tau)]=S \mathbf{C}[\rho,\mathcal{A}_{\text{L}}, \mathcal{A}_{\text{MAI}}(\tau)] R^T$ and $\boldsymbol{\Gamma}[\rho,\mathbf{X}(\tau)]=S \boldsymbol{\Gamma}[\rho,\mathcal{A}_{\text{MAI}}(\tau)] S^T$, the moment matrix in Eq.~\eqref{eq:momentmatrix} can be reexpressed as
\begin{align}
    \mathbf{M}[\rho,\mathbf{G},\mathbf{X}(\tau)] &:= \mathbf{C}[\rho,\mathbf{G},\mathbf{X}(\tau)]^T \boldsymbol{\Gamma}[\rho,\mathbf{X}(\tau)]^{-1} \mathbf{C}[\rho,\mathbf{G},\mathbf{X}(\tau)] \nonumber \\
    &= R \mathbf{C}[\rho,\mathcal{A}_{\text{L}}, \mathcal{A}_{\text{MAI}}(\tau)]^T S^T \left( S \boldsymbol{\Gamma}[\rho,\mathcal{A}_{\text{MAI}}(\tau)] S^T \right)^{-1} S \mathbf{C}[\rho,\mathcal{A}_{\text{L}}, \mathcal{A}_{\text{MAI}}(\tau)] R^T \nonumber \\
    & \leq R \mathbf{M}[\rho,\mathcal{A}_{\text{L}}, \mathcal{A}_{\text{MAI}}(\tau)] R^T,
\end{align}
where we define the moment matrix $\mathbf{M}[\rho,\mathcal{A}_{\text{L}}, \mathcal{A}_{\text{MAI}}(\tau)]=\mathbf{C}[\rho,\mathcal{A}_{\text{L}}, \mathcal{A}_{\text{MAI}}(\tau)]^T \boldsymbol{\Gamma}[\rho,\mathcal{A}_{\text{MAI}}(\tau)]^{-1} \mathbf{C}[\rho,\mathcal{A}_{\text{L}}, \mathcal{A}_{\text{MAI}}(\tau)]$. Here the inequality in the last line derived from the matrix-valued Cauchy-Schwarz inequality~\cite{ManuelNC2020}. It is saturated if the measurement $\mathbf{X}(\tau)$ is optimized over the accessible vector $\mathcal{A}_{\text{MAI}}(\tau)$, i.e. $\max_{\mathbf{X}(\tau)\in \text{span}(\mathcal{A}_{\text{MAI}}(\tau))} \mathbf{M}[\rho,\mathbf{G},\mathbf{X}(\tau)] = R \mathbf{M}[\rho,\mathcal{A}_{\text{L}}, \mathcal{A}_{\text{MAI}}(\tau)] R^T$. For any given generator matrix $R$, the matrix $S$ corresponding to optimal measurements is given as $GS=R \mathbf{C}[\rho,\mathcal{A}_{\text{L}}, \mathcal{A}_{\text{MAI}}(\tau)]^T \boldsymbol{\Gamma}[\rho,\mathcal{A}_{\text{MAI}}(\tau)] ^{-1}$, with a real matrix $G$  for normalization.

For a give generator vector $\mathbf{G}$, the squeezing matrix optimizing over accessible measurements $\mathcal{A}_{\text{MAI}}(\tau)$ can be written in terms of the moment matrix, 
\begin{align}
\Xi^2_{\text{opt}}[\rho,\mathbf{G},\mathcal{A}_{\text{MAI}}(\tau)] &= \min_{\mathbf{X}(\tau)\in\text{span}(\mathcal{A}_{\text{MAI}}(\tau))} \Xi^2[\rho,\mathbf{G},\mathbf{X}(\tau)] \nonumber \\
&= \min_{\mathbf{X}(\tau)\in\text{span}(\mathcal{A}_{\text{MAI}}(\tau))} \mathbf{F}_{\text{SN}}[\mathbf{G}]^\frac{1}{2} \mathbf{M}[\rho,\mathbf{G},\mathbf{X}]^{-1} \mathbf{F}_{\text{SN}}[\mathbf{G}]^{\frac{1}{2} } \nonumber \\
&= \mathbf{F}_{\text{SN}}[\mathbf{G}]^\frac{1}{2} R \mathbf{M}[\rho,\mathcal{A}_{\text{L}}, \mathcal{A}_{\text{MAI}}(\tau)]^{-1} R^T \mathbf{F}_{\text{SN}}[\mathbf{G}]^{\frac{1}{2} },
\end{align}
which yields the corresponding covariance matrix
\begin{align}
\Sigma[\rho,\mathbf{G},\mathcal{A}_{\text{MAI}}(\tau)] = \Sigma_{\text{SN}}^{\frac{1}{2}} \Xi^2_{\text{opt}}[\rho,\mathbf{G},\mathcal{A}_{\text{MAI}}(\tau)] \Sigma_{\text{SN}}^{\frac{1}{2}}.
\end{align}
After optimizing $R$ and the evolution times $\tau$, the covariance matrix can be further optimized as $\Sigma=\min_{R,\tau} \Sigma[\rho,\mathbf{G},\mathcal{A}_{\text{MAI}}(\tau)]$. For a given linear combination $\mathbf{n}$, one can finally obtain the relative estimation uncertainty as
\begin{align}
\xi^{-2}(\mathbf{n}) = \frac{\mathbf{n}^T \Sigma_{\text{SN}} \mathbf{n} }{\mathbf{n}^T \Sigma \mathbf{n}}.
\end{align}
We assume the vector $\mathbf{n}=\left(n_1,\cdots,n_M \right)$ containing equally distributed elements with different signs, i.e. $n_m=\pm \frac{1}{\sqrt{M}}$. In our scenarios where the phase generators are locally implemented, we find that the resulting multiparameter sensitivity is independent with the array of signs: $\xi^{-2}(\mathbf{n})=\xi^{-2}$.

In the atomic ensembles considered in the main text, since the squeezing of the OAT spin states mostly lie in the $yz-$plane, the vector of linear operators on each mode is set as $\mathcal{L}^{(m)}=\left(S^{(m)}_y,S^{(m)}_z \right)$, where $S^{(m)}_\beta=\sum_i^{N_m}\sigma^{(i)}_\beta/2$ is local collective spin observables for $N_m$ atoms in the $m$-th mode. The additional state evolution in nonlocal and local MAI protocols can be described by the unitary operator $U_\text{nl}(\tau)=e^{-i \chi \tau \left( \sum_{m=1}^M S_{z}^{(m)}\right)^2 }$ and $U^{(m)}_\text{loc}(\tau)=e^{-i \chi \tau  \left(S_{z}^{(m)} \right)^2}$, respectively, which yields the corresponding vectors $\mathbf{X}_{\text{MAI},\alpha}^{(m)}=\left( U_{\alpha}^\dagger (\tau) S^{(m)}_y U_{\alpha} (\tau), U_{\alpha}^\dagger (\tau) S^{(m)}_z U_{\alpha} (\tau) \right)$. Then, based on $\mathcal{L}^{(m)}$ and $\mathbf{X}_{\text{MAI},\alpha}^{(m)}$, we can obtain the family of accessible linear and MAI measurements on $M$-mode $\mathcal{A}_{\text{L}}$ and $\mathcal{A}_{\text{MAI},\alpha}$. 

To simplify the model, the atom numbers in every sub-ensembles are assumed to be the same, i.e. $N_m=N/M$. In this case, the $2M\times 2M$ covariance matrix can be expressed in terms of 2 submatrices $\boldsymbol{\Gamma}_{mm}$ and $\boldsymbol{\Gamma}_{mn}$,
\begin{align}\label{eq:Gamma2M}
\boldsymbol{\Gamma} = \left( \begin{matrix}
\boldsymbol{\Gamma}_{mm} & \boldsymbol{\Gamma}_{mn} & \cdots & \boldsymbol{\Gamma}_{mn} \\
\boldsymbol{\Gamma}_{mn} & \boldsymbol{\Gamma}_{mm} & \cdots & \boldsymbol{\Gamma}_{mn} \\
\vdots & \vdots  & \ddots & \vdots \\
\boldsymbol{\Gamma}_{mn} & \boldsymbol{\Gamma}_{mn} & \cdots & \boldsymbol{\Gamma}_{mm}
\end{matrix}\right),
\end{align}
where $\boldsymbol{\Gamma}_{mm}$ is a $2\times 2$ covariance matrix of local operators, whose elements are $\left( \Gamma_{mm} \right)_{ij}=\text{Cov}\left( \mathcal{A}_{\text{MAI},\alpha,i}^{(m)}, \mathcal{A}_{\text{MAI},\alpha,j}^{(m)}\right)$, $\boldsymbol{\Gamma}_{mn}$ is a $2\times 2$ covariance matrix of nonlocal operators, with elements $\left( \Gamma_{mn} \right)_{ij}=\text{Cov}\left( \mathcal{A}_{\text{MAI},\alpha,i}^{(m)}, \mathcal{A}_{\text{MAI},\alpha,j}^{(n)}\right) (m\neq n)$. For mode separable pure states $\rho_{\text{MS}}$, the submatrices of nonlocal operators become zero matrix $\boldsymbol{\Gamma}_{mn}=\mathbf{0}_{2\times 2}$, thus the covariance matrix can be expressed as $\boldsymbol{\Gamma}=\oplus_{m=1}^M \boldsymbol{\Gamma}_{mm}$.

Similarly, the $2M \times 2M$ commutator matrix $\boldsymbol{C}$  can be expressed in terms of submatrices $\boldsymbol{C}_{mm}$ and $\boldsymbol{C}_{mn}$ 
\begin{align}\label{eq:Commutator2M}
\boldsymbol{C} = \left( \begin{matrix}
\boldsymbol{C}_{mm} & \boldsymbol{C}_{mn} & \cdots & \boldsymbol{C}_{mn} \\
\boldsymbol{C}_{mn} & \boldsymbol{C}_{mm} & \cdots & \boldsymbol{C}_{mn} \\
\vdots & \vdots  & \ddots & \vdots \\
\boldsymbol{C}_{mn} & \boldsymbol{C}_{mn} & \cdots & \boldsymbol{C}_{mm}
\end{matrix}\right),
\end{align}
where $\boldsymbol{C}_{mm}$ is a $2\times 2$ commutator matrix of local operators, with elements $\left( C_{mm} \right)_{ij}=i\langle [ \mathcal{L}_i^{(m)}, \mathbf{X}_{\text{MAI},\alpha,j}^{(m)} ]\rangle$, and $\boldsymbol{C}_{mn}$ is a $2\times 2$ commutator matrix of nonlocal operators, with elements $\left( C_{mn} \right)_{ij}=i\langle [ \mathcal{L}_i^{(m)}, \mathbf{X}_{\text{MAI},\alpha,j}^{(n)} ]\rangle$. In the local MAI protocols, the submatrices $\boldsymbol{C}_{mn}=\mathbf{0}_{2\times 2}$ will lead to $\boldsymbol{C}=\oplus_{m=1}^M \boldsymbol{C}_{mm}$. Based on $\boldsymbol{\Gamma}$ and $\boldsymbol{C}$, the moment matrix $\mathbf{M}=\boldsymbol{C}^T\boldsymbol{\Gamma}\boldsymbol{C}$ can be constructed.

To optimize the phase generators matrix $R$, we express the local vectors $\mathbf{g}_m$ in terms of measurement angles $\phi_m$:  $\mathbf{g}_m=\left( \cos{(\phi_m)}, \sin{(\phi_m)} \right)$. Therefore, the minimized squeezing matrix is
\begin{align}
\min \Xi^2_{\text{opt}}[\rho,\mathbf{G},\mathcal{A}_{\text{MAI}}(\tau)]= \min_{\tau \in \text{span} T} \min_{\phi_1,\cdots,\phi_M}  \mathbf{F}_{\text{SN}}[\mathbf{G}]^\frac{1}{2} R \mathbf{M}[\rho,\mathcal{A}_{\text{L}}, \mathcal{A}_{\text{MAI}}(\tau)]^{-1} R^T \mathbf{F}_{\text{SN}}[\mathbf{G}]^{\frac{1}{2} }, 
\end{align}
where $T$ is accessible time range and $\mathbf{F}_{\text{SN}}=\text{diag}(N_1,\cdots,N_M)$. Because the metrological sensitivity is independent with the arrays of signs in $\mathbf{n}$, we can set $\mathbf{n}=\mathbf{n}_+ =\frac{1}{\sqrt{M}}(1,\cdots,1)$ without loss of generality, in which case the optimal local directions for $R$ are the same, i.e. $\phi_m=\phi$.

\section{II.\quad Analytical solutions for Multiparameter moment matrix for spin squeezed states}

\subsection{A. local squeezing spin states $\rho_{\text{MS}}$}
We consider an ensemble of $N$ spin-$1/2$ particles initially polarized along the $x$ direction are distributed among $M$ spatially-separated modes, and assume the atom number in each mode is the same, i.e. $N_m=N/M=\mathcal{N}$. To generate the local squeezing spin states, the probe state will undergo the OAT interaction described by the local Hamiltonian
\begin{align}
\tilde{H}_{\text{loc}}  &=\hbar \chi \sum_{m=1}^{M} \left( S_{z}^{(m)} \right)^2,
\end{align}
which can generate the particle entanglement locally in each modes. The resulting mode separable states are written as $\rho_{\text{MS}}=|\psi_{\text{MS}} \rangle \langle \psi_{\text{MS}}|$, where $\psi_{\text{MS}}=e^{-i  \frac{\tilde{H}_{\text{loc}}}{\hbar}t  } |\psi(0)\rangle$. In the following, we denote the OAT squeezing times for state preparation as $\mu=2\chi t$, and denote the OAT evolution times for MAI strategies as $\mu_{\alpha}=2\chi \tau$.

\subsection{Multiparameter moment matrix for $\rho_{\text{MS}}$ in typical protocols using linear measurements}
For mode separable pure states $\rho_{\text{MS}}$, the covariance matrix can be expressed by $\boldsymbol{\Gamma} = \oplus_{m=1}^M \boldsymbol{\Gamma}_{mm}$, where the submatrix $\boldsymbol{\Gamma}_{mm}$ is given as
\begin{align}
\boldsymbol{\Gamma}_{mm} &= \left( \begin{matrix}
\text{Cov}\left( S_y^{(m)}, S_y^{(m)} \right) & \text{Cov}\left( S_y^{(m)}, S_z^{(m)} \right) \\
\text{Cov}\left( S_z^{(m)}, S_y^{(m)} \right) & \text{Cov}\left( S_z^{(m)}, S_z^{(m)} \right)
\end{matrix} \right) , \\
\text{Cov} \left( S^{(m)}_y, S^{(m)}_y \right) &= \frac{1}{8} \mathcal{N} \left( \mathcal{N}+1-(\mathcal{N}-1)\cos{(\mu)}^{\mathcal{N}-2} \right), \\
\text{Cov} \left( S^{(m)}_y, S^{(m)}_z \right) &= \text{Cov} \left( S^{(m)}_z, S^{(m)}_y \right) = \frac{1}{4} \mathcal{N}(\mathcal{N}-1)\cos{\left( \frac{\mu}{2} \right)}^{\mathcal{N}-2} \sin{\left( \frac{\mu}{2} \right)}, \\
\text{Cov} \left( S^{(m)}_z, S^{(m)}_z \right) &= \frac{\mathcal{N}}{4}.
\end{align}
The commutator matrix is $\boldsymbol{C}=\oplus_{m=1}^M \boldsymbol{C}_{mm}$, where
\begin{align}
\boldsymbol{C}_{mm} &= i \left( \begin{matrix} 
0 & \langle [S_y^{(m)}, S_z^{(m)} ]\rangle \\
\langle [S_z^{(m)}, S_y^{(m)} ] & 0
\end{matrix} \right), \\
-i\langle [S_y^{(m)}, S_z^{(m)} ]\rangle &= i\langle [S_z^{(m)}, S_y^{(m)} ]\rangle = \frac{1}{2} \mathcal{N} \cos{\left( \frac{\mu}{2} \right)}^{\mathcal{N}-1}.
\end{align}

\subsection{Multiparameter moment matrix for $\rho_{\text{MS}}$ in local MAI protocols}
The $2M \times 2M$ covariance matrix $\boldsymbol{\Gamma}$ can be written as $\boldsymbol{\Gamma}=\oplus_{m=1}^M \boldsymbol{\Gamma}_{mm}$, where
\begin{align}
\boldsymbol{\Gamma}_{mm} &= \left( \begin{matrix}
\text{Cov}\left( U_{\text{loc}}^\dagger S_y^{(m)} U_{\text{loc}},  U_{\text{loc}}^\dagger S_y^{(m)} U_{\text{loc}} \right) &  \text{Cov}\left( U_{\text{loc}}^\dagger S_y^{(m)} U_{\text{loc}},  U_{\text{loc}}^\dagger S_z^{(m)} U_{\text{loc}} \right) \\
\text{Cov}\left( U_{\text{loc}}^\dagger S_z^{(m)} U_{\text{loc}},  U_{\text{loc}}^\dagger S_y^{(m)} U_{\text{loc}} \right) & \text{Cov}\left( U_{\text{loc}}^\dagger S_z^{(m)} U_{\text{loc}},  U_{\text{loc}}^\dagger S_z^{(m)} U_{\text{loc}} \right)
\end{matrix} \right) , \\
\text{Cov}\left( U_{\text{loc}}^\dagger S_y^{(m)} U_{\text{loc}},  U_{\text{loc}}^\dagger S_y^{(m)} U_{\text{loc}} \right) &= \frac{1}{8} \mathcal{N} \left( \mathcal{N}+1-(\mathcal{N}-1)\cos{(\mu-\mu_{\text{loc}})}^{\mathcal{N}-2} \right), \\
\text{Cov}\left( U_{\text{loc}}^\dagger S_y^{(m)} U_{\text{loc}},  U_{\text{loc}}^\dagger S_z^{(m)} U_{\text{loc}} \right) &= \text{Cov}\left( U_{\text{loc}}^\dagger S_z^{(m)} U_{\text{loc}},  U_{\text{loc}}^\dagger S_y^{(m)} U_{\text{loc}} \right) = \frac{1}{4} \mathcal{N}(\mathcal{N}-1)\cos{\left( \frac{\mu-\mu_{\text{loc}}}{2} \right)}^{\mathcal{N}-2} \sin{\left( \frac{\mu-\mu_{\text{loc}}}{2} \right)}, \\
\text{Cov}\left( U_{\text{loc}}^\dagger S_z^{(m)} U_{\text{loc}},  U_{\text{loc}}^\dagger S_z^{(m)} U_{\text{loc}} \right) &= \frac{\mathcal{N}}{4}.
\end{align}
The commutator matrix is $\boldsymbol{C}=\oplus_{m=1}^M \boldsymbol{C}_{mm}$, where
\begin{align}
\boldsymbol{C}_{mm} &= i \left( \begin{matrix}
\langle [S_y^{(m)}, U_{\text{loc}}^\dagger S_y^{(m)} U_{\text{loc}} ] \rangle & \langle [S_y^{(m)}, U_{\text{loc}}^\dagger S_z^{(m)} U_{\text{loc}} ] \rangle \\
\langle [S_z^{(m)}, U_{\text{loc}}^\dagger S_y^{(m)} U_{\text{loc}} ] \rangle & \langle [S_z^{(m)}, U_{\text{loc}}^\dagger S_z^{(m)} U_{\text{loc}} ] \rangle
\end{matrix} \right),\\
i \langle [S_y^{(m)}, U_{\text{loc}}^\dagger S_y^{(m)} U_{\text{loc}} ] \rangle &= -\frac{1}{4} \mathcal{N}(\mathcal{N}-1) \left( \cos{\left( \mu-\frac{\mu_{\text{loc}}}{2} \right)}^{\mathcal{N}-2} +\cos{\left( \frac{\mu_{\text{loc}}}{2} \right)}^{\mathcal{N}-2}   \right)\sin{\left( \frac{\mu_{\text{loc}}}{2} \right)}, \\
i \langle [S_y^{(m)}, U_{\text{loc}}^\dagger S_z^{(m)} U_{\text{loc}} ] \rangle &= \frac{1}{2} \mathcal{N} \cos{ \left( \frac{\mu}{2} \right)}^{\mathcal{N}-1} , \\
i \langle [S_z^{(m)}, U_{\text{loc}}^\dagger S_y^{(m)} U_{\text{loc}} ] \rangle &=  -\frac{1}{2} \mathcal{N} \cos{\left(\frac{\mu-\mu_{\text{loc}}}{2} \right)}^{\mathcal{N}-1}, \\
i \langle [S_z^{(m)}, U_{\text{loc}}^\dagger S_z^{(m)} U_{\text{loc}} ] \rangle &= 0.
\end{align}

\subsection{B. nonlocal squeezing spin states  $\rho_{\text{ME}}$}
To generate the nonlocal squeezing spin states, we consider the nonlocal OAT interaction
\begin{align}
\tilde{H}_{\text{nl}}  &=\hbar \chi \left( \sum_{m=1}^{M}  S_{z}^{(m)} \right)^2.
\end{align}
This Hamiltonian can simultaneously generate mode entanglement and particle entanglement. The resulting mode entangled states are written as $\rho_{\text{ME}}=|\psi_{\text{ME}} \rangle \langle \psi_{\text{ME}}|$, where $\psi_{\text{ME}}=e^{-i  \frac{\tilde{H}_{\text{nl}}}{\hbar}t  } |\psi(0)\rangle$.

\subsection{Multiparameter moment matrix for $\rho_{\text{ME}}$ in typical protocols using linear measurements}
For mode entangled state $\rho_{\text{ME}}$, the covariance matrix is given as Eq.~\eqref{eq:Gamma2M}, where the covariance matrix $\boldsymbol{\Gamma}_{mm}$ is
\begin{align}
\boldsymbol{\Gamma}_{mm} &= \left( \begin{matrix}
\text{Cov}\left( S_y^{(m)}, S_y^{(m)} \right) & \text{Cov}\left( S_y^{(m)}, S_z^{(m)} \right) \\
\text{Cov}\left( S_z^{(m)}, S_y^{(m)} \right) & \text{Cov}\left( S_z^{(m)}, S_z^{(m)} \right)
\end{matrix} \right) , \\
\text{Cov} \left( S^{(m)}_y, S^{(m)}_y \right) &= \frac{1}{8} \mathcal{N} \left( \mathcal{N}+1-(\mathcal{N}-1)\cos{(\mu)}^{M \mathcal{N}-2} \right), \\
\text{Cov} \left( S^{(m)}_y, S^{(m)}_z \right) &= \text{Cov} \left( S^{(m)}_z, S^{(m)}_y \right) = \frac{1}{4} \mathcal{N}(\mathcal{N}-1)\cos{\left( \frac{\mu}{2} \right)}^{M \mathcal{N}-2} \sin{\left( \frac{\mu}{2} \right)}, \\
\text{Cov} \left( S^{(m)}_z, S^{(m)}_z \right) &= \frac{\mathcal{N}}{4},
\end{align}
and the covariance matrix $\boldsymbol{\Gamma}_{mn}$ is
\begin{align}
\boldsymbol{\Gamma}_{mn} &= \left( \begin{matrix}
\text{Cov}\left( S_y^{(m)}, S_y^{(n)} \right) & \text{Cov}\left( S_y^{(m)}, S_z^{(n)} \right) \\
\text{Cov}\left( S_z^{(m)}, S_y^{(n)} \right) & \text{Cov}\left( S_z^{(m)}, S_z^{(n)} \right)
\end{matrix} \right) , \\
\text{Cov} \left( S^{(m)}_y, S^{(n)}_y \right) &= \frac{1}{8} \mathcal{N}^2 \left(1-\cos{(\mu)}^{M \mathcal{N}-2} \right), \\
\text{Cov} \left( S^{(m)}_y, S^{(n)}_z \right) &= \text{Cov} \left( S^{(m)}_z, S^{(n)}_y \right) = \frac{1}{4} \mathcal{N}^2 \cos{\left( \frac{\mu}{2} \right)}^{M \mathcal{N}-2} \sin{\left( \frac{\mu}{2} \right)} , \\
\text{Cov} \left( S^{(m)}_z, S^{(n)}_z \right) &= 0.
\end{align}

In the typical protocol, the commutator matrix can be expressed as $\boldsymbol{C} = \oplus_{m=1}^M \boldsymbol{C}_{mm}$, where $\boldsymbol{C}_{mm}$ is given as
\begin{align}
\boldsymbol{C}_{mm} &= i \left( \begin{matrix} 
0 & \langle [S_y^{(m)}, S_z^{(m)} ]\rangle \\
\langle [S_z^{(m)}, S_y^{(m)} ] & 0
\end{matrix} \right), \\
-i\langle [S_y^{(m)}, S_z^{(m)} ]\rangle &= i\langle [S_z^{(m)}, S_y^{(m)} ]\rangle = \frac{1}{2} \mathcal{N} \cos{\left( \frac{\mu}{2} \right)}^{M \mathcal{N}-1} 
\end{align}

\subsection{Multiparameter moment matrix for $\rho_{\text{ME}}$ in nonlocal MAI protocols}
The $2M \times 2M$ covariance matrix $\boldsymbol{\Gamma}$ can be expressed in terms of the submatrices $\boldsymbol{\Gamma}_{mm}, \boldsymbol{\Gamma}_{mn}$ as Eq.~\eqref{eq:Gamma2M}, where
\begin{align}
\boldsymbol{\Gamma}_{mm} &= \left( \begin{matrix}
\text{Cov}\left( U_{\text{nl}}^\dagger S_y^{(m)} U_{\text{nl}},  U_{\text{nl}}^\dagger S_y^{(m)} U_{\text{nl}} \right) &  \text{Cov}\left( U_{\text{nl}}^\dagger S_y^{(m)} U_{\text{nl}},  U_{\text{nl}}^\dagger S_z^{(m)} U_{\text{nl}} \right) \\
\text{Cov}\left( U_{\text{nl}}^\dagger S_z^{(m)} U_{\text{nl}},  U_{\text{nl}}^\dagger S_y^{(m)} U_{\text{nl}} \right) & \text{Cov}\left( U_{\text{nl}}^\dagger S_z^{(m)} U_{\text{nl}},  U_{\text{nl}}^\dagger S_z^{(m)} U_{\text{nl}} \right)
\end{matrix} \right) , \\
\text{Cov}\left( U_{\text{nl}}^\dagger S_y^{(m)} U_{\text{nl}},  U_{\text{nl}}^\dagger S_y^{(m)} U_{\text{nl}} \right) &= \frac{1}{8} \mathcal{N} \left( \mathcal{N}+1-(\mathcal{N}-1)\cos{(\mu-\mu_{\text{nl}})}^{M \mathcal{N}-2} \right), \\
\text{Cov}\left( U_{\text{nl}}^\dagger S_y^{(k)} U_{\text{nl}},  U_{\text{nl}}^\dagger S_z^{(m)} U_{\text{nl}} \right) &= \text{Cov}\left( U_{\text{nl}}^\dagger S_z^{(m)} U_{\text{nl}},  U_{\text{nl}}^\dagger S_y^{(m)} U_{\text{nl}} \right) = \frac{1}{4} \mathcal{N}(\mathcal{N}-1)\cos{\left( \frac{\mu-\mu_{\text{nl}}}{2} \right)}^{M \mathcal{N}-2} \sin{\left( \frac{\mu-\mu_{\text{nl}}}{2} \right)}, \\
\text{Cov}\left( U_{\text{nl}}^\dagger S_z^{(m)} U_{\text{nl}},  U_{\text{nl}}^\dagger S_z^{(m)} U_{\text{nl}} \right) &= \frac{\mathcal{N}}{4}, 
\end{align}
and 
\begin{align}
\Gamma_{mn} &= \left( \begin{matrix}
\text{Cov}\left( U_{\text{nl}}^\dagger S_y^{(m)} U_{\text{nl}},  U_{\text{nl}}^\dagger S_y^{(n)} U_{\text{nl}} \right) &  \text{Cov}\left( U_{\text{nl}}^\dagger S_y^{(m)} U_{\text{nl}},  U_{\text{nl}}^\dagger S_z^{(n)} U_{\text{nl}} \right) \\
\text{Cov}\left( U_{\text{nl}}^\dagger S_z^{(m)} U_{\text{nl}},  U_{\text{nl}}^\dagger S_y^{(n)} U_{\text{nl}} \right) & \text{Cov}\left( U_{\text{nl}}^\dagger S_z^{(m)} U_{\text{nl}},  U_{\text{nl}}^\dagger S_z^{(n)} U_{\text{nl}} \right)
\end{matrix} \right) , \\
\text{Cov}\left( U_{\text{nl}}^\dagger S_y^{(m)} U_{\text{nl}},  U_{\text{nl}}^\dagger S_y^{(n)} U_{\text{nl}} \right) &= \frac{1}{8} \mathcal{N}^2 \left(1-\cos{(\mu-\mu_{\text{nl}})}^{M \mathcal{N}-2} \right), \\
\text{Cov}\left( U_{\text{nl}}^\dagger S_y^{(m)} U_{\text{nl}},  U_{\text{nl}}^\dagger S_z^{(n)} U_{\text{nl}} \right) &= \text{Cov}\left( U_{\text{nl}}^\dagger S_z^{(m)} U_{\text{nl}},  U_{\text{nl}}^\dagger S_y^{(n)} U_{\text{nl}} \right) = \frac{1}{4} \mathcal{N}^2 \cos{\left( \frac{\mu-\mu_{\text{nl}}}{2} \right)}^{M \mathcal{N}-2} \sin{\left( \frac{\mu-\mu_{\text{nl}}}{2} \right)} , \\
\text{Cov}\left( U_{\text{nl}}^\dagger S_z^{(m)} U_{\text{nl}},  U_{\text{nl}}^\dagger S_z^{(n)} U_{\text{nl}} \right) &= 0.
\end{align}

The commutator matrix $\boldsymbol{C}$ can be expressed by $\boldsymbol{C}_{mm}$ and $\boldsymbol{C}_{mn}$, where $\boldsymbol{C}_{mm}$ is a $2\times 2$ submatrix with elements $\left( \boldsymbol{C}_{mm} \right)_{ij}=i\langle [ \mathcal{L}_i^{(m)}, \mathbf{X}_{\text{MAI,nl},j}^{(m)} ]\rangle$,
\begin{align}
\boldsymbol{C}_{mm} &= i \left( \begin{matrix}
\langle [S_y^{(m)}, U_{\text{nl}}^\dagger S_y^{(m)} U_{\text{nl}} ] \rangle & \langle [S_y^{(m)}, U_{\text{nl}}^\dagger S_z^{(m)} U_{\text{nl}} ] \rangle \\
\langle [S_z^{(m)}, U_{\text{nl}}^\dagger S_y^{(m)} U_{\text{nl}} ] \rangle & \langle [S_z^{(m)}, U_{\text{nl}}^\dagger S_z^{(m)} U_{\text{nl}} ] \rangle
\end{matrix} \right),\\
i \langle [S_y^{(m)}, U_{\text{nl}}^\dagger S_y^{(m)} U_{\text{nl}} ] \rangle &= -\frac{1}{4} \mathcal{N}(\mathcal{N}-1) \left( \cos{\left( \mu-\frac{\mu_{\text{nl}}}{2} \right)}^{M \mathcal{N}-2} +\cos{\left( \frac{\mu_{\text{nl}}}{2} \right)}^{M \mathcal{N}-2}   \right)\sin{\left( \frac{\mu_{\text{nl}}}{2} \right)}, \\
i \langle [S_y^{(m)}, U_{\text{nl}}^\dagger S_z^{(m)} U_{\text{nl}} ] \rangle &= \frac{1}{2} \mathcal{N} \cos{ \left( \frac{\mu}{2} \right)}^{M \mathcal{N}-1} , \\
i \langle [S_z^{(m)}, U_{\text{nl}}^\dagger S_y^{(m)} U_{\text{nl}} ] \rangle &=  -\frac{1}{2} \mathcal{N} \cos{\left(\frac{\mu-\mu_{\text{nl}}}{2} \right)}^{M \mathcal{N}-1}, \\
i \langle [S_z^{(m)}, U_{\text{nl}}^\dagger S_z^{(m)} U_{\text{nl}} ] \rangle &= 0.
\end{align}
The other submatrix $\boldsymbol{C}_{mn}$ whose elements are $\left(\boldsymbol{C}_{mn} \right)_{ij}=i\langle [ \mathcal{L}_i^{(m)}, \mathbf{X}_{\text{MAI,nl},j}^{(n)} ]\rangle $ is expressed as
\begin{align}
\boldsymbol{C}_{mn} &= i \left( \begin{matrix}
\langle [S_y^{(m)}, U_{\text{nl}}^\dagger S_y^{(n)} U_{\text{nl}} ] \rangle & \langle [S_y^{(m)}, U_{\text{nl}}^\dagger S_z^{(n)} U_{\text{nl}} ] \rangle \\
\langle [S_z^{(m)}, U_{\text{nl}}^\dagger S_y^{(n)} U_{\text{nl}} ] \rangle & \langle [S_z^{(m)}, U_{\text{nl}}^\dagger S_z^{(n)} U_{\text{nl}} ] \rangle
\end{matrix}\right), \\
i \langle [S_y^{(m)}, U_{\text{nl}}^\dagger S_y^{(n)} U_{\text{nl}} ] \rangle &= -\frac{1}{4} \mathcal{N}^2 \left( \cos{\left( \mu-\frac{\mu_{\text{nl}}}{2} \right)}^{M \mathcal{N}-2} +\cos{\left( \frac{\mu_{\text{nl}}}{2} \right)}^{M \mathcal{N}-2}   \right)\sin{\left( \frac{\mu_{\text{nl}}}{2} \right)}, \\
\langle [S_y^{(m)}, U_{\text{nl}}^\dagger S_z^{(n)} U_{\text{nl}} ] \rangle &= i \langle [S_z^{(m)}, U_{\text{nl}}^\dagger S_y^{(n)} U_{\text{nl}} ] \rangle = \langle [S_z^{(m)}, U_{\text{nl}}^\dagger S_z^{(n)} U_{\text{nl}} ] \rangle =0.
\end{align}

\subsection{Multiparameter moment matrix for $\rho_{\text{ME}}$ in local MAI protocols}
As in Eq.~\eqref{eq:Gamma2M}, the $2M \times 2M$ covariance matrix $\boldsymbol{\Gamma}$ can be expressed by two submatrices $\boldsymbol{\Gamma}_{mm}, \boldsymbol{\Gamma}_{mn}$, which are expressed as
\begin{align}
& \boldsymbol{\Gamma}_{mm} = \left( \begin{matrix}
\text{Cov}\left( (U_{\text{loc}}^{(m)})^\dagger S_y^{(m)} U_{\text{loc}}^{(m)},  (U_{\text{loc}}^{(m)})^\dagger S_y^{(m)} U_{\text{loc}}^{(m)} \right) &  \text{Cov}\left( (U_{\text{loc}}^{(m)})^\dagger S_y^{(m)} U_{\text{loc}}^{(m)},  (U_{\text{loc}}^{(m)})^\dagger S_z^{(m)} U_{\text{loc}}^{(m)} \right) \\
\text{Cov}\left( (U_{\text{loc}}^{(m)})^\dagger S_z^{(m)} U_{\text{loc}}^{(m)},  (U_{\text{loc}}^{(m)})^\dagger S_y^{(m)} U_{\text{loc}}^{(m)} \right) & \text{Cov}\left( (U_{\text{loc}}^{(m)})^\dagger S_z^{(m)} U_{\text{loc}}^{(m)},  (U_{\text{loc}}^{(m)})^\dagger S_z^{(m)} U_{\text{loc}}^{(m)} \right)
\end{matrix} \right) , \\
& \text{Cov}\left( (U_{\text{loc}}^{(m)})^\dagger S_y^{(m)} U_{\text{loc}}^{(m)}, (U_{\text{loc}}^{(m)})^\dagger S_y^{(m)} U_{\text{loc}}^{(m)} \right) = 
\frac{1}{8} \mathcal{N} \left(1+\mathcal{N}-(-1+\mathcal{N})\cos{(\mu)}^{(M-1)\mathcal{N}} \cos{(\mu-\mu_{\text{loc}})}^{\mathcal{N}-2} \right) , \\
& \text{Cov}\left( (U_{\text{loc}}^{(m)})^\dagger S_y^{(m)} U_{\text{loc}}^{(m)},  (U_{\text{loc}}^{(m)})^\dagger S_z^{(m)} U_{\text{loc}}^{(m)} \right) = \text{Cov}\left( (U_{\text{loc}}^{(m)})^\dagger S_z^{(m)} U_{\text{loc}}^{(m)},  (U_{\text{loc}}^{(m)})^\dagger S_y^{(m)} U_{\text{loc}}^{(m)} \right) \nonumber \\
&\; = \frac{1}{4}\mathcal{N}(\mathcal{N}-1) \cos{\left( \frac{\mu}{2} \right)}^{(M-1)\mathcal{N}}\cos{\left( \frac{\mu-\mu_{\text{loc}}}{2} \right)}^{\mathcal{N}-2} \sin{\left( \frac{\mu-\mu_{\text{loc}}}{2} \right)} , \\
& \text{Cov}\left( (U_{\text{loc}}^{(m)})^\dagger S_z^{(m)} U_{\text{loc}}^{(m)},  (U_{\text{loc}}^{(m)})^\dagger S_z^{(m)} U_{\text{loc}}^{(m)}  \right) = \frac{\mathcal{N}}{4}, 
\end{align}
and 
\begin{align}
& \boldsymbol{\Gamma}_{mn} = \left( \begin{matrix}
\text{Cov}\left( (U_{\text{loc}}^{(m)})^\dagger S_y^{(m)} U_{\text{loc}}^{(m)},  (U_{\text{loc}}^{(n)})^\dagger S_y^{(n)} U_{\text{loc}}^{(n)} \right) &  \text{Cov}\left( (U_{\text{loc}}^{(m)})^\dagger S_y^{(m)} U_{\text{loc}}^{(m)},  (U_{\text{loc}}^{(n)})^\dagger S_z^{(n)} U_{\text{loc}}^{(n)} \right) \\
\text{Cov}\left( (U_{\text{loc}}^{(m)})^\dagger S_z^{(m)} U_{\text{loc}}^{(m)},  (U_{\text{loc}}^{(n)})^\dagger S_y^{(n)} U_{\text{loc}}^{(n)} \right) & \text{Cov}\left( (U_{\text{loc}}^{(m)})^\dagger S_z^{(m)} U_{\text{loc}}^{(m)},  (U_{\text{loc}}^{(n)})^\dagger S_z^{(n)} U_{\text{loc}}^{(n)} \right)
\end{matrix} \right) , \\
& \text{Cov}\left( (U_{\text{loc}}^{(m)})^\dagger S_y^{(m)} U_{\text{loc}}^{(m)},  (U_{\text{loc}}^{(n)})^\dagger S_y^{(n)} U_{\text{loc}}^{(n)} \right) = \frac{1}{8} \mathcal{N}^2 \left( -\cos{(\mu)}^{(M-2)\mathcal{N}} \cos{\left( \mu-\frac{\mu_{\text{loc}}}{2} \right)}^{2\mathcal{N}-2}+\cos{\left( \frac{\mu_{\text{loc}}}{2} \right)}^{2\mathcal{N}-2}  \right), \\
& \text{Cov}\left( (U_{\text{loc}}^{(m)})^\dagger S_y^{(m)} U_{\text{loc}}^{(m)},  (U_{\text{loc}}^{(n)})^\dagger S_z^{(n)} U_{\text{loc}}^{(n)} \right) = \text{Cov}\left( (U_{\text{loc}}^{(m)})^\dagger S_z^{(m)} U_{\text{loc}}^{(m)},  (U_{\text{loc}}^{(n)})^\dagger S_y^{(n)} U_{\text{loc}}^{(n)} \right) \nonumber \\
&\;= \frac{1}{4}\mathcal{N}^2 \cos{\left( \frac{\mu}{2} \right)}^{(M-1)\mathcal{N}-1} \cos{\left( \frac{\mu-\mu_{\text{loc}}}{2} \right)}^{\mathcal{N}-1} \sin{\left( \frac{\mu}{2} \right)} , \\
&\text{Cov}\left( (U_{\text{loc}}^{(m)})^\dagger S_z^{(m)} U_{\text{loc}}^{(m)},  (U_{\text{loc}}^{(n)})^\dagger S_z^{(n)} U_{\text{loc}}^{(n)} \right) = 0.
\end{align}

\begin{figure}[t]
    \begin{center}
	\includegraphics[width=140mm]{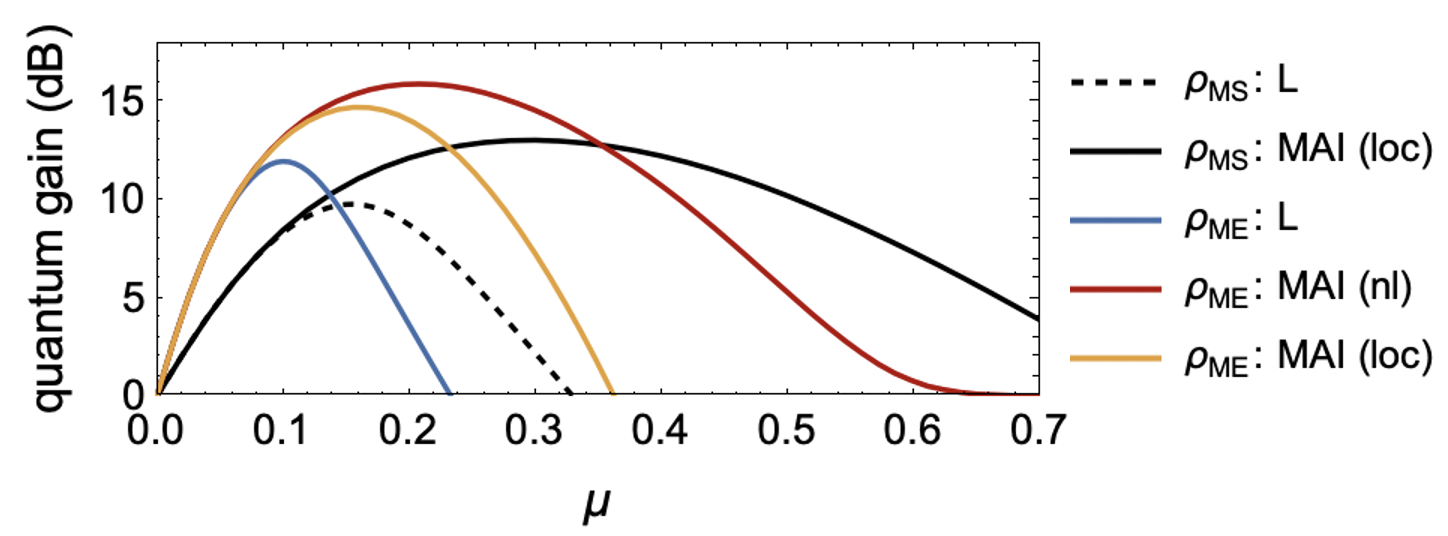}
	\end{center}
    \caption{ Comparison of multiparameter squeezing for mode-separable states $\rho_{\text{MS}}$ and mode-entangled states $\rho_{\text{ME}}$ when the total atom number $N=100$ and the mode number $M=2$. It is observed that under the same measurement strategies, $\rho_{\text{MS}}$ can achieve a higher sensitivity. It indicates that the entanglement between the modes leads to higher metrological sensitivity.}
    \label{SMFig_spinMSME}
\end{figure}

The $2M\times 2M$ commutator matrix is $\boldsymbol{C} = \oplus_{m=1}^M \boldsymbol{C}_{mm}$, where $\boldsymbol{C}_{mm}$ is given as
\begin{align}
\boldsymbol{C}_{mm} &= i \left( \begin{matrix} 
\langle [S_y^{(m)}, (U_{\text{loc}}^{(m)})^\dagger S_y^{(m)} U_{\text{loc}}^{(m)} ]\rangle & \langle [S_y^{(m)}, (U_{\text{loc}}^{(m)})^\dagger S_z^{(m)} U_{\text{loc}}^{(m)} ]\rangle \\
\langle [S_z^{(m)}, (U_{\text{loc}}^{(m)})^\dagger S_y^{(m)} U_{\text{loc}}^{(m)} ]\rangle & \langle [S_y^{(m)}, (U_{\text{loc}}^{(m)})^\dagger S_z^{(m)} U_{\text{loc}}^{(m)} ]\rangle
\end{matrix} \right), \\
i\langle [S_y^{(m)}, (U_{\text{loc}}^{(m)})^\dagger S_y^{(m)} U_{\text{loc}}^{(m)} ]\rangle &=  \frac{1}{4}\mathcal{N}(\mathcal{N}-1)\sin{\left( \frac{\mu_{\text{loc}}}{2} \right)} \left( \cos{(\mu)}^{(M-1)\mathcal{N}} \cos{\left( \mu-\frac{\mu_{\text{loc}}}{2} \right)}^{\mathcal{N}-2} + \cos{\left( \frac{\mu_{\text{loc}}}{2} \right)}^{\mathcal{N}-2} \right) , \\
i\langle [S_y^{(m)}, (U_{\text{loc}}^{(m)})^\dagger S_z^{(k)} U_{loc}^{(k)} ]\rangle &= -\frac{1}{2} \mathcal{N}\cos{\left( \frac{\mu}{2} \right)}^{M\mathcal{N}-1} , \\
i\langle [S_z^{(m)}, (U_{\text{loc}}^{(m)})^\dagger S_y^{(m)} U_{\text{loc}}^{(m)} ]\rangle &=  \frac{1}{2} \mathcal{N} \cos{\left( \frac{\mu}{2} \right)}^{(M-1)\mathcal{N}} \cos{\left( \frac{\mu-\mu_{\text{loc}}}{2} \right)}^{\mathcal{N}-1}, \\
i\langle [S_z^{(m)}, (U_{\text{loc}}^{(m)})^\dagger S_z^{(m)} U_{\text{loc}}^{(m)} ]\rangle &= 0.
\end{align}

In Fig.~\ref{SMFig_spinMSME}, we compare the multiparameter squeezing detected from mode-separable states $\rho_{\text{MS}}$ and mode-entangled states $\rho_{\text{ME}}$. It is observed that the existence of mode entanglement contributes to achieving a higher sensitivity.

\section{III.\quad Multiparameter squeezing for two-mode squeezed states }
A two mode squeezed vacuum state can be prepared by a two mode squeezing operator $S_{\text{AB}}=e^{-\zeta a^\dagger b^\dagger+\zeta^* ab} $, with the squeezing parameter $\zeta=r e^{i\phi}$. Without loss of generality, we assume $\phi=0$. A complete family of linear operator is made of $\mathcal{L}=\left(x,p\right)$, where $x=(a+a^\dagger)/\sqrt{2}$ and $p=-i(a-a^\dagger)/\sqrt{2}$ are quadrature observables. Therefore, the family of linear measurements over two-mode reads $\mathcal{A}_{\text{L}}=\left( x_A,p_A,x_B,p_B\right)$. In the typical scenario where only linear measurements are considered, the covariance and commutator matrix are
\begin{align}
\boldsymbol{\Gamma} &= \frac{1}{2} \left( \begin{matrix}
\cosh{(2r)} & 0 & -\sinh{(2r)} & 0 \\
0 & \cosh{(2r)} & 0 & \sinh{(2r)} \\
-\sinh{(2r)} & 0 & \cosh{(2r)} & 0 \\
0 & \sinh{(2r)} & 0 & \cosh{(2r)}
\end{matrix} \right) , \\
\boldsymbol{C} &= \left( \begin{matrix}
0 & 1 & 0 & 0 \\
-1 & 0 & 0 & 0 \\
0 & 0 & 0 & 1 \\
0 & 0 & -1 & 0
\end{matrix} \right).
\end{align}
After optimizing the measurements, we obtain the maximum sensitivity 
\begin{align}\label{eq:deltaL}
    \xi^{-2}_{\text{L}} = e^{2r}.
\end{align}

Then, we investigate the MAI strategies, where an additional evolution $U_{\text{MAI}}$ is introduced before linear measurements. We first consider a nonlocal evolution, which is performed by a two-mode squeezing operator $U_{\text{nl}}=e^{r_{\text{nl}}(a^\dagger b^\dagger -a b)} $. Based on the associated family $\mathcal{A}_{\text{MAI}}=\left( U_{\text{nl}}^\dagger x_A U_{\text{nl}}, U_{\text{nl}}^\dagger p_A U_{\text{nl}}, U_{\text{nl}}^\dagger x_B U_{\text{nl}}, U_{\text{nl}}^\dagger p_B U_{\text{nl}}\right)$, the covariance and commutator matrices can be written as
\begin{align}
\boldsymbol{\Gamma} &= \frac{1}{2} \left( \begin{matrix}
\cosh{(2(r-r_{\text{nl}}))} & 0 & -\sinh{(2(r-r_{\text{nl}}))} & 0 \\
0 & \cosh{(2(r-r_{\text{nl}}))} & 0 & \sinh{(2(r-r_{\text{nl}}))} \\
-\sinh{(2(r-r_{\text{nl}}))} & 0 & \cosh{(2(r-r_{\text{nl}}))} & 0 \\
0 & \sinh{(2(r-r_{\text{nl}}))} & 0 & \cosh{(2(r-r_{\text{nl}}))}
\end{matrix} \right),\\
\boldsymbol{C} &= \left( \begin{matrix} 
0 & \cosh{(r_{\text{nl}})} & 0 & -\sinh{(r_{\text{nl}})} \\
-\cosh{(r_{\text{nl}})} & 0 & -\sinh{(r_{\text{nl}})} & 0 \\
0 & -\sinh{(r_{\text{nl}})} & 0 & \cosh{(r_{\text{nl}})} \\
-\sinh{(r_{\text{nl}})} & 0 & -\cosh{(r_{\text{nl}})} & 0 
\end{matrix} \right).
\end{align}
The maximum sensitivity in the nonlocal MAI strategy is 
\begin{align}\label{eq:deltaMAInl}
\xi^{-2}_{\text{MAI,nl}} = e^{2r},
\end{align}
which is independent with the MAI-dependent squeezing parameter $r_{\text{nl}}$.

In local MAI strategies, local evolutions performed by one-mode squeezing operators $U_{\text{loc}}^{(1)}=S_A=e^{ r_{\text{loc}}( a^2 - a^{\dagger 2} )/2}, U_{\text{loc}}^{(2)}=S_B=e^{ r_{\text{loc}} ( b^2 -b^{\dagger 2} )/2}$ are considered. The vector of MAI operators reads $\mathcal{A}_{\text{MAI}}=\left( S_A^\dagger x_A S_A, S_A^\dagger p_A S_A, S_B^\dagger x_B S_B, S_B^\dagger p_B S_B \right)$, and the resulting covariance and commutator matrix are
\begin{align}
\boldsymbol{\Gamma} &= \frac{1}{2} \left( \begin{matrix} 
e^{2 r_{ \text{loc} }} \cosh{(2r)} & 0 & -e^{2r_{ \text{loc} }}\sinh{(2r)} & 0 \\
0 & e^{-2 r_{ \text{loc} }} \cosh{(2r)} & 0 & e^{-2r_{ \text{loc} }}\sinh{(2r)} \\
-e^{2r_{ \text{loc} }}\sinh{(2r)} & 0 & e^{2 r_{ \text{loc} }} \cosh{(2r)} & 0 \\
0 & e^{-2 r_{ \text{loc} }} \sinh{(2r)} & 0 & e^{-2r_{ \text{loc} }}\cosh{(2r)}
\end{matrix}\right) , \\
\boldsymbol{C} &=\left( \begin{matrix} 
0 & e^{-r_{ \text{loc} }} & 0 & 0 \\
-e^{r_{ \text{loc} }} & 0 & 0 & 0 \\
0 & 0 & 0 & e^{-r_{ \text{loc} }} \\
0 & 0 & -e^{r_{ \text{loc} }} & 0
\end{matrix} \right).
\end{align}
The maximum sensitivity after the measurement optimization is
\begin{align}\label{eq:deltaMAIloc}
\xi^{-2}_{\text{MAI,loc}} = e^{2r}.
\end{align}
It is observed that the sensitivities in Eqs.~\eqref{eq:deltaL}\eqref{eq:deltaMAInl}\eqref{eq:deltaMAIloc} are the same, which solely depend on the initial squeezing parameter $r$. 

To show the advantage from MAI technique in a realistic scenario, we consider the effect of detection noise. In this case, the measurement is written as $\tilde{x}^\alpha=x^\alpha+\Delta x^\alpha$ and $\tilde{p}^\alpha=p^\alpha+\Delta p^\alpha$, where $\Delta X$ is a random variable following the Gaussian distribution with mean value $\langle \Delta X \rangle =0 $ and variance $\langle (\Delta X)^2 \rangle=\sigma^2$. Influenced by detection noise, the sensitivity from the typical protocol becomes
\begin{align}
\xi^{-2}_{\text{L}} &= \frac{1}{e^{-2r}+2\sigma^2}.
\end{align}
In MAI protocols, the MAI operator under the detection noise is $U_{\alpha}^\dagger \tilde{X} U_{\alpha}=U_{\alpha}^\dagger X U_{\alpha}+\Delta X$. We obtain the sensitivity from the MAI protocols with nonlocal and local squeezing operators as
\begin{align}
\xi^{-2}_{\text{MAI,nl}} &= \frac{1}{e^{-2r}+2e^{-2 r_{\text{nl}} } \sigma^2} , \\
\xi^{-2}_{\text{MAI,loc}} &= \frac{1}{e^{-2r}+2e^{-2 r_{ \text{loc} } } \sigma^2}.
\end{align}
We demonstrate that the MAI protocols show improve noise robustness as long as $r_\alpha>0$.

\end{widetext}

\end{document}